\documentclass[twocolumn,floats,prl,showpacs]{revtex4}

\usepackage{amsmath}
\usepackage{amsfonts}

\usepackage{graphicx}

\newcommand {\be}{\begin{equation}}
\newcommand {\ee} {\end{equation}}
\newcommand {\bea}{\begin{eqnarray}}
\newcommand {\eea} {\end{eqnarray}}

\newcommand{\bk}{{\bf k}}

\newcommand{\kom}{\left| \Omega \right>}

\newcommand{\koma}{\left| \Omega_A \right>}
\newcommand{\komb}{\left| \Omega_B \right>}
\newcommand{\ket}[1]{\left| #1 \right>}

\begin{document}


\title{Entanglement spectra of complex paired superfluids}
\author{J. Dubail and N. Read}
\affiliation{Department of Physics, Yale
University, P.O. Box 208120, New Haven, CT 06520-8120, USA}
\date{August 15, 2011}

\begin{abstract}
We study the entanglement in various fully-gapped complex paired states of fermions in two dimensions, focusing on the entanglement spectrum (ES), and using the Bardeen-Cooper-Schrieffer (BCS) form of the ground state wavefunction on a cylinder. Certain forms of the pairing functions allow a simple and explicit exact solution for the ES. In the weak-pairing phase of $\ell$-wave paired spinless fermions ($\ell$ odd), the universal low-lying part of the ES consists of $|\ell|$ chiral Majorana fermion modes [or $2|\ell|$ ($\ell$ even) for spin-singlet states]. For $|\ell|>1$, the pseudo-energies of the modes are split in general, but for all $\ell$ there is a zero--pseudo-energy mode at zero wavevector if the number of modes is odd. This ES agrees with the perturbed conformal field theory of the edge excitations. For more general BCS states, we show how the entanglement gap diverges as a model pairing function is approached.
\end{abstract}

\pacs{74.20.Fg, 74.20.Rp, 03.67.Mn}

\maketitle


The hunt for both theoretical and experimental methods
that would allow to distinguish between two different topological phases of matter \cite{toporder,MR} has
been recently revived, greatly due to the question whether the experimentally-observed fractional quantum Hall effect (FQHE) at $\nu =5/2$ supports excitations with non-Abelian statistics \cite{MR,stern}. An experimental observation of these non-Abelian anyons, possibly in interferometry experiments \cite{interferometry1,interferometry2,interferometry3,interferometry4}, would be a major breakthrough and an important step forward in the field of topological quantum computation \cite{reviewTQC}. Other proposals for experimental realizations of non-Abelian phases of matter include two-dimensional (2D) $p$-wave superfluids \cite{rg}, such as films of some phase of $^3$He \cite{Volovik}, ultracold Fermi gases (atoms interacting via $p$-wave Feschbach resonance \cite{Gurarie} or micro-wave dressed polar molecules \cite{CooperShlyap}), or closely related physics at
the interface between a superconductor and a topological insulator \cite{FuKane}.

On the theoretical side, our understanding of the collective behavior
of electrons and their emergent properties (such as properties of the excitations, including fractional charge and statistics) relies mostly on trial wave-functions \cite{BCS,Laughlin}. Numerical comparisons between the trial wave-functions and the realistic ones ({\it e.g.\/} for Coulomb interactions) are most meaningful when they focus on quantities that are robust inside a topological phase. In recent years, quantum information concepts \cite{nielsenchuang,QI} such as quantum entanglement have provided valuable insights in this matter. First, the topological entanglement entropy (EE) associated with a reduced density matrix of a 2D ground state in a topological phase \cite{KitaevPreskill,LevinWen} was shown to take a constant value throughout the phase. Two different phases, however, might have the same topological EE. More refined approaches consider the full structure of the reduced density matrix. Its set of eigenvalues was studied numerically by Li and Haldane (LH) \cite{HaldaneLi}. They defined the {\it entanglement spectrum} (ES) as the set of real {\it pseudo-energies} $\varepsilon_n$ appearing in the Schmidt decomposition of the ground state $\ket{\psi}$ for a (spatial) bipartition of the Hilbert space $\mathcal{H} = \mathcal{H}_A \otimes \mathcal{H}_B$:
\begin{equation}
	\ket{\psi} = \frac{1}{\sqrt{\mathcal{Z}}} \sum_{n} e^{- \varepsilon_n/2} \ket{\psi_{A,n}} \otimes \ket{\psi_{B,n}}
\end{equation}
with $\mathcal{Z} = \sum_n e^{- \varepsilon_n}$.
For the FQHE at $\nu =5/2$, LH argued that as the size goes to infinity, the ES contains a universal low--pseudo-energy part ({\it i.e.\/}, separated by a gap from the rest of the ES), which is related to the physical edge spectrum of this phase \cite{milr} (such a relation with a chiral conformal field theory was also anticipated in Ref.\ \cite{KitaevPreskill}). This entanglement gap varies inside the topological phase; it goes to infinity for a particular model wave function: the Moore-Read (MR) wave function \cite{MR}.

The ES has been studied in various systems, including FQHE
\cite{ESFQHE,ESFQHE2}, spin chains or ladders \cite{thomaleXXZ} and lattice complex paired superfluids \cite{EShaas}.
Most of the results rely on numerical calculations of the ES. Exact analytical results in basic examples or toy models would shed light on this topic. However, such progress has been largely precluded by
the intrinsic technical difficulty of studying these highly entangled phases of matter.

In this Letter we study 2D complex paired superfluids \cite{rg}, starting from BCS theory \cite{BCS}, and show that the ES can be worked out explicitly for certain model wave functions. The problem essentially reduces to a free fermion problem, for which various other methods are known \cite{EislerPeschel}. To the best of our knowledge, these tools
have not been used so far to make analytic calculations of the ES in topological superfluids, however, there are general arguments about the relation with the edge spectrum \cite{Fidkowski,tzv,qkl}.


{\it Complex paired superfluids---} In the BCS theory \cite{BCS} for translation-invariant systems, the fermions in the ground state form pairs, the members of which carry opposite momenta $\bk$ and $-\bk$. The mean-field Hamiltonian for spinless or spin-polarized particles is
\begin{equation}
\label{eq:BCSham}
H_{{\rm BCS}} = \sum_\bk \left[\xi_\bk c^\dagger_\bk c_\bk + \frac{1}{2} \left( \overline{\Delta}_\bk c_{-\bk} c_{\bk} + \Delta_\bk c_{\bk}^\dagger c_{-\bk}^\dagger \right) \right],
\end{equation}
where $\xi_\bk = \frac{k^2}{2 m^*} - \mu$ is the kinetic energy minus the chemical potential for each single particle, $\Delta_\bk$ is the gap function ($\overline{\Delta}_\bk$ its complex conjugate), and $c_\bk^\dagger, c_\bk$ are fermion creation/annihilation modes. The ground-state of (\ref{eq:BCSham}) is
\begin{equation}
	\label{eq:groundstate}
	\kom \; \propto\; \exp \left( \frac{1}{2} \sum_{\bk} g_{\bk} c^\dagger_{-\bk} c^\dagger_\bk \right) \ket{0},
\end{equation}
$g_{\bk} = \left(\xi_\bk - \sqrt{\xi_\bk^2+|\Delta_\bk|^2}\right)/\overline{\Delta}_\bk$ is the {\it pairing function}.

Initially we will assume that, as a function of $\bk$, $\Delta_\bk$ is an eigenfunction of rotations, with angular momentum $\ell$ (the relative angular momentum of the particles in a pair). The Fermi statistics of the particles allows only odd values of $\ell$; in such states, parity and time-reversal symmetry are broken.  We first consider $\ell=-1$ ($\ell=+1$ is similar). In two dimensions, the $\ell = -1$ gap function behaves generically as $\Delta_\bk \simeq \widehat{\Delta} (k_x - i k_y)$ at small $\bk$ ($\widehat{\Delta}>0$ without loss of generality). At small $\bk$, the behavior of $g_\bk$ depends on the sign of
$\mu$, and there is a topological phase transition at $\mu=0$ \cite{rg}. In the {\it strong-pairing phase} ($\mu <0$), $g_\bk \sim (k_x -i k_y)$ as $\bk\to 0$, and the pairing function in real space decays exponentially with the distance between the particles in a pair. At large scale, the fluid can be thought of as a Bose condensate of point-like Cooper pairs. In the {\it weak-pairing phase} ($\mu >0$), $g_\bk \sim 1/(k_x+ i k_y)$, and the pairing function decays algebraically. At large distances, the component of the ground state (\ref{eq:groundstate}) with $N$ particles at complex coordinates $z_j=x_j+iy_j$ ($j=1$, \ldots, $N$) is given by \cite{rg}
\begin{equation}
	\label{eq:pfaffian}
	\psi(z_1,\dots,z_N) = \left<0 \right| c_{z_1} \dots c_{z_N} \kom \sim  {\rm Pf} \left\{ \frac{1}{z_i -z_j}\right\},
\end{equation}
where ${\rm Pf}$ denotes the Pfaffian of a matrix.
This Pfaffian form is a factor in the MR wave function \cite{MR}, and its long-range behavior is the first indication of the topologically non-trivial nature of the weak-pairing $p\pm ip$ phase \cite{rg}.

\begin{figure}[htbp]
\includegraphics[width=0.42\textwidth]{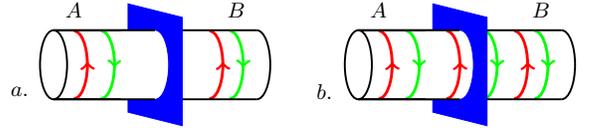}
\caption{(color online) Origin of the two terms in the Schmidt decomposition (\ref{eq:Schmidtp}). (a) Independent configurations of pairs in $A$ and $B$. (b) Shared Cooper pair between $A$ and $B$.}
\label{fig:shared}
\end{figure}

{\it Schmidt decomposition on the cylinder--- } We consider the BCS wave function
(\ref{eq:groundstate}) on an infinite cylinder of circumference $L_y$, with coordinates $(x,y) \in \mathbb{R} \times \left[0,L_y \right)$. Rotational invariance of the cylinder implies that the ground state is a product:
\begin{equation}
	\label{eq:tensorproduct}
	\kom = \bigotimes_{k_y \geq 0} \kom_{k_y},
\end{equation}
where the $\pm k_y$'s are the discrete $y$-components of momentum for the particles ($k_y L_y/2 \pi$ is either integer or half-integer depending on the boundary conditions of the fermions). The Schmidt decomposition for a transverse cut of the cylinder at $x=0$ (Fig. \ref{fig:shared}) is obtained as follows. Let us call $A$ (resp. $B$) the semi-infinite cylinder with $x \leq 0$ (resp. $x>0$). First, one introduces the (normalized) {\it cut ground state} $\ket{\Omega_A}_{k_y} \otimes \ket{\Omega_B}_{k_y}$ for each $k_y \geq 0$, with
\begin{equation}
	\label{eq:komkyA}
	\ket{\Omega_A}_{k_y} \, \propto \,e^{ \int_{-\infty}^0 dx_+ dx_- \, g_{k_y}(x_-,x_+) c_{x_- ,-k_y}^\dagger c_{x_+,k_y}^\dagger } \ket{0},
\end{equation}
where all the particles have transverse momentum $\pm k_y $ and are within part
$A$ ($x \leq 0$), and a similar definition holds for $\ket{\Omega_B}_{k_y}$ with $x>0$.
Then, the (normalized) wave function $\kom_{k_y}$ is created from the cut ground state
\begin{equation}
	\label{eq:komky}
	\ket{\Omega}_{k_y} \, \propto  \, e^{G^\dagger_{k_y}} \, \koma_{k_y}\otimes \komb_{k_y},
\end{equation}
with a {\it sharing operator} $G^\dagger_{k_y}$ generating {\it shared pairs} (one particle in $A$, one in $B$)
\begin{equation}
	\label{eq:sharingop}
	G_{k_y}^\dagger = \int_{-\infty}^\infty d x_+\, \int_{-\infty}^\infty d x_-\, g_{k_y}^{AB}(x_-,x_+) c_{x_-,-k_y}^\dagger c^\dagger_{x_+,k_y},
\end{equation}
where $g^{AB}_{k_y}(x_-,x_+) = g_{k_y}(x_-,x_+)$ if $(x_-,x_+) \in A \times B$ or $(x_-,x_+) \in B \times A$, and $0$ otherwise. $G_{k_y}^\dagger$ (i.e.\ $g_{k_y}$) can be Schmidt decomposed
\begin{equation}
	\label{eq:Gdagger}
	G_{k_y}^\dagger  = \sum_{n=1}^r \alpha_n(k_y) d^\dagger_{A,k_y,n} d^\dagger_{B,k_y ,n},
\end{equation}
where $d^\dagger_{A,k_y,n}$ ($d^\dagger_{B,k_y ,n}$) creates a particle in one of an orthonormal set of single-particle states lying entirely in $A$ (resp.,\ $B$),
$r$ is the Schmidt rank of $G^\dagger_{k_y}$, and $\alpha_n(k_y)$ are some coefficients. Expanding the exponential in (\ref{eq:komky}), one thus gets $2^r$ terms. Generically, $2^r$ is the size of the smaller Hilbert space $\mathcal{H}_A$ or $\mathcal{H}_B$, and is infinite in our continuum BCS model, so it is difficult to make progress from eq.\ (\ref{eq:komky}). Moreover, the different terms in the expansion are not orthogonal in general.

{\it Model BCS wave functions---} However, for one trial form an exact solution is possible. This is the case when $\Delta_\bk=\widehat{\Delta}(k_x-ik_y)$ for {\em all} $\bk$, and $\mu= m^* \widehat{\Delta}^2/2$, then
\begin{equation}
	\label{eq:pairingp}
	g_\bk = \frac{- m^*\widehat{\Delta}}{k_x + i k_y}
\end{equation}
for all $\bk$, and in the plane $g(x,y)\propto 1/z$ for {\em all} $x$, $y$. (This special point has been studied in Ref.\ \cite{Sierra}.) For a finite density of particles, an ultraviolet cutoff is needed, which for the cylinder we take to be a cutoff at large $k_y$.
The pairing function in terms of $k_y>0$ and $x_\pm$, eq.\ (\ref{eq:pairingp}), is
\begin{equation}
	\label{eq:pairing1dp}
	g_{k_y}(x_-,x_+)\, =\, i \, m^* \widehat{\Delta} \, e^{-k_y (x_--x_+)} \, \Theta(x_--x_+),
\end{equation}
where $\Theta(x)$ is the Heaviside step function, and so the shared pairing function $g_{k_y}^{AB}(x_-,x_+) \propto e^{- k_y x_-} e^{k_y x_+}$ is factorized ($x_+$ must be in $A$ and $x_-$ in $B$). This leads to a Schmidt decomposition for $G_{k_y}^\dagger$ of rank $r=1$:
\begin{eqnarray}
	\label{eq:analytic1}
	G_{k_y}^\dagger &=& \frac{\mu/\widehat{\Delta}}{k_y} d^\dagger_{k_y,A} d^\dagger_{k_y,B}, \\
\nonumber	d^\dagger_{A,k_y} &=& e^{-i \frac{\pi}{4}} \int_{-\infty}^0 dx \sqrt{2k_y} e^{k_y x} c_{x,k_y}^\dagger,\\
\nonumber	d^\dagger_{B,k_y} &=& e^{-i \frac{\pi}{4}} \int^{\infty}_0 dx \sqrt{2k_y} e^{-k_y x} c_{x,-k_y}^\dagger.
\end{eqnarray}
Thus, expanding the exponential in (\ref{eq:komky}), we get only two terms.
The first term corresponds to configurations with independent Cooper pairs in $A$ and $B$, and the second one to configurations with one Cooper pair shared between $A$ and $B$ (Fig. \ref{fig:shared}). The two terms are obviously orthogonal, because (for $k_y\neq0$) they have
different total transverse momentum in $A$ (or $B$).
Their norms can be computed using Wick's theorem and two-point correlation functions in the original ground state $\kom$: $\left< d_A d_A^\dagger \right> = \left< d_B d_B^\dagger \right> = v/\sqrt{1+v^2}$
and $\left< d_A^\dagger d_B^\dagger \right> = v \left( 1-v /\sqrt{1+v^2} \right)$, where $v=k_y/(m^* \widehat{\Delta})$. The Schmidt decomposition of $\kom_{k_y}$ becomes
\begin{eqnarray}
	\label{eq:Schmidtp}
	\kom_{k_y} &=& \frac{1}{\sqrt{1+e^{-\varepsilon(k_y)}}} \left[ \koma_{k_y} \otimes \komb_{k_y} \right. \\
\nonumber	 &&\left.  + \; e^{-\varepsilon(k_y)/2} \,b_{A,k_y}^\dagger \koma_{k_y} \otimes b_{B,k_y}^\dagger \komb_{k_y} \right],
\end{eqnarray}
with $b^\dagger_{A,k_y} \koma \propto d^\dagger_{A,k_y} \koma$ and $b^\dagger_{B,k_y} \koma \propto d^\dagger_{B,k_y} \koma$, normalized such that $\left< \Omega_A \right| b_{A,k_y} b_{A,k_y}^\dagger \koma = \left< \Omega_B \right| b_{B,k_y} b_{B,k_y}^\dagger \komb =1$, and the pseudo-energy is
\begin{equation}
	\label{eq:pseudoenergy}
	\varepsilon(k_y) = 2 \ln \left( \sqrt{1+(k_y /m^*\widehat{\Delta})^2 }+ k_y/m^* \widehat{\Delta}  \right).
\end{equation}

{\it Chiral ES---} Thus, the ES of $\kom$ for the model state consists of a fermion mode of pseudo-energy $\varepsilon(k_y)$ for each $k_y\geq0$ appearing in (\ref{eq:tensorproduct}). If we introduce fermion operators $\eta_{k}^\dagger=\eta_{-k}$ for all $k=\pm k_y$, $\left\{ \eta_{k} ,\eta_{k'} \right\} = \delta_{k+k',0}$, then the spectrum of the pseudo-Hamiltonian
\begin{equation}
	\label{eq:spectrum}
	H_{ES} \,= \, \sum_{k_y \geq 0} \varepsilon(k_y) \,\eta_{-k_y} \eta_{k_y}
\end{equation}
reproduces the ES. At small $k_y$, $\varepsilon(k_y) \simeq 2 k_y/(m^* \widehat{\Delta})$, so the ES is that of a chiral Majorana fermion theory.
In the case of periodic boundary condition in the $y$-direction, $k_y=0$ is included in the product
(\ref{eq:tensorproduct}). The expressions in eq.\ (\ref{eq:analytic1}) are divergent, and the $x$ integrals should be cutoff at $x=\pm L_x/2$. There are again two terms in the Schmidt decomposition of $\kom_{k_y=0}$, and $\varepsilon(0)=0$. We have incorporated this above by introducing a self-adjoint (``Majorana'') operator $\eta_0^\dagger=\eta_0$ (which requires a two-dimensional space of states for $k_y=0$). Another such operator (located far away along the cylinder) would be needed to make a complex fermion operator (similar to the case of the edge spectrum on a cylinder \cite{milr}).


{\it Higher angular momenta---} The generalization to paired states with general $\ell<0$---modeled by $\Delta_\bk = \widehat{\Delta} (k_x-i k_y)^{|\ell|}$ in the Hamiltonian (\ref{eq:BCSham})---reveals an interesting structure. For odd angular momenta $\ell$ in the weak-pairing phase $\mu>0$, we consider the model pairing function $g_{\bk} = -2\mu/[\widehat{\Delta}(k_x+i k_y)^{|\ell|}]$ for spinless fermions.  For even $\ell$ we consider the spin-singlet case with the same form of model pairing function (appearing in the wave function as $g_{\bk} c_{-\bk \downarrow}^\dagger c_{\bk \uparrow}^\dagger$). In contrast to the $\ell=-1$ case, there is no value of $\mu$ for which these are the exact functions, but they still capture the correct long-distance behavior of the weak-pairing phase. We find (for fixed spin orientation for even $\ell$)
\begin{equation}
	\nonumber
	g_{k_y}(x_{-},x_+)  \, \propto  \, (x_--x_+)^{|\ell|-1} e^{-k_y(x_--x_+)} \Theta(x_--x_+).
\end{equation}
The sharing operator (\ref{eq:Gdagger}) has rank $r=|\ell|$, leading to a Schmidt decomposition of $\kom_{k_y}$ of rank $2^{|\ell|}$. The calculation of the pseudo-energies can be done as in the $p$-wave case using Wick's theorem, although this now involves diagonalizing a matrix of rank $|\ell|$. For example, for
$\ell = -2$, recombining the two spin sectors ($\uparrow$ or $\downarrow$ for particles
with momentum $+k_y$), we find that the ES at small $k_y$ corresponds to
\begin{equation}
\nonumber	H_{ES} = \sum_{k_y \geq 0} \left[ \frac{\sqrt{2}\widehat{\Delta}k_y}{\mu} \eta_{a ,s, -k_y} \eta_{a, s,k_y} + M_{ab} \eta_{a,s,-k_y} \eta_{b,s,k_y}  \right]
\end{equation}
with sums over repeated indices $a$, $b=1$, $2$ and $s = \uparrow$, $\downarrow$. Here $M = m \sigma_y$ is a splitting matrix \cite{rCFTedge}, and $m = 2 \ln(1+\sqrt{2})$. The pseudo-energy eigenvalues are then given by $\varepsilon_\pm(k_y)=\pm m+(\sqrt{2}\widehat{\Delta}k_y/\mu)$ and are spin-degenerate. We expect a similar form in general, with $|\ell|$ chiral Majorana fields ($2|\ell|$ with spin) and an $|\ell|\times|\ell|$ antisymmetric Hermitian splitting matrix $M$ (the term linear in $k_y$ might contain different velocities for different fermion types). Thus when $|\ell|$ is even, the modes at $k_y\to 0$ come in pairs with pseudo-energy $\pm m_\alpha$, $\alpha=1$, \ldots, $|\ell|/2$ (with spin degeneracy). When $|\ell|$ is odd, $M$ has at least one zero eigenvalue, and generically we are again left with one Majorana zero mode operator at $k_y=0$ {\em only}. This is consistent with arguments that these weak-pairing phases of $\ell$-wave superfluids should support non-Abelian vortex excitations when $\ell$ is odd, but not when $\ell$ is even \cite{rg}. (Another case is spin-triplet odd-$\ell$ pairing, which for model pairing functions like those above leads to $2|\ell|$ modes, with $\ell$ odd \cite{rg}.)


We remark that the model wave functions we have used possess an interesting property, that is directly related to the number of Majorana fields which appear in the ES. The configurations with non-zero amplitude in the wave function of $\kom_{k_y}$ ($k_y>0$) obey some specific rules. For $|\ell| = 1$, the fermions with $+k_y$ and with $-k_y$ must alternate in $x$ (Fig.\ \ref{fig:shared}). Similarly, for $|\ell| = 2$, there can be at most {\it two} successive fermions with identical momentum-spin $+k_y\uparrow$, but {\it not three}, and the same for $-k_y\downarrow$ fermions. For general $\ell$, configurations with up to $|\ell|$ successive $+k_y$ (resp., $-k_y$) momenta can occur, but not more.
Similarly to the $p$-wave case (Fig. \ref{fig:shared}), there are several possibilities for the configurations of particles across the cut between $A$ and $B$, leading to the $2^{|\ell|}$ (or $2^{2|\ell|}$) terms in the Schmidt decomposition (\ref{eq:Schmidtp}).


{\it Entanglement gap---} In the weak-pairing phase, for $\mu \neq m^* \widehat{\Delta}^2/2$, the pairing function can be written $g_\bk = f(|\bk|^2) g_{\bk}^{(0)}$
with $f(|\bk|^2) \rightarrow 1$ when $|\bk| \rightarrow 0$, and $g_\bk^{(0)}$ is the model pairing function (\ref{eq:pairingp}) (we consider $\ell=-1$ for definiteness). For fixed $k_y \geq 0$, there is a new length scale $r_0(k_y) \equiv \sqrt{|f'(k_y^2)|}$ in the one-dimensional wave function $\kom_{k_y}$. The pairing function in real $x$-space $g_{k_y}$ is a convolution of $g_{k_y}^{(0)}$ ($\ref{eq:pairing1dp}$) with a kernel of width $\sim r_0(k_y)$. In other words, for $|x_--x_+| \gg r_0(k_y)$, one has $g_{k_y}(x_-,x_+) \simeq g_{k_y}^{(0)}(x_-,x_+)$, but the behavior of the pairing function is different for $|x_--x_+| \lesssim r_0(k_y)$. This produces (infinitely many) additional terms in the decomposition of the corresponding sharing operator $G_{k_y}$
The additional terms in the Schmidt decomposition come from configurations with pairs of
particles with momenta $\pm k_y$ in the region close to the cut $-r_0(k_y) \lesssim x_- <0$, $ 0< x_+ \lesssim r_0(k_y)$ (or with $x_-$ and $x_+$ exchanged).
Close to the special point $\mu = m^* \widehat{\Delta}^2/2$, $r_0(k_y)$ is small, and one expects
such terms to be of order $\sim r_0(k_y)^2 \left< \rho_{\pm k_y}(-r_0) \rho_{\mp k_y}(r_0) \right>$
where $\rho_{k_y}(x) = c^\dagger_{x,k_y} c_{x,k_y}$ is the density of particles with momentum
$k_y$. Apart from a small change in the chiral mode
(\ref{eq:pseudoenergy}) (which must still
behave linearly when $k_y\rightarrow 0$, with a coefficient depending on $\mu$), the only effect is that additional pseudo-energies appear in the ES, of order $  -\ln \left[ r_0(k_y)^2 \left< \rho_{\pm k_y}(-r_0) \rho_{\mp k_y}(r_0) \right> \right]$. At $k_y=0$ we get a rough estimate of the gap $\sim -\ln \left| 1 -\mu/(m^* \widehat{\Delta}^2/2) \right|$.

{\it Strong-pairing phase---} In the strong-pairing phase,
the exponential decay of the pairing function may be approximated by (derivatives of) a delta function, e.g.\ $g_{k_y}(x) \propto (\partial_x + k_y) \delta(x)$ for $\ell=-1$. The term $k_y \delta(x)$ is local, so it does not generate any entanglement. The term $\delta'(x)$ might give some non-trivial contribution to the ES, but one can repeat similar arguments as for the entanglement gap. A regularization of $\delta'(x)$ would be localized on a short length $r_0$. For a finite density of particles $\left<\rho(k_y)\right>$, one would then find that all the pseudo-energies go to infinity as $r_0 \rightarrow 0$. We conclude that, provided we keep the density of particles $\left< \rho(k_y) \right>$ finite, the ES in the strong-pairing phase cannot contain a
universal gapless part. This is expected, because the strong-pairing
phase is adiabatically connected to the vacuum as $\mu\to-\infty$ \cite{rg}. Note that for s-wave pairing, $\ell=0$, there is only the strong-pairing phase.


{\it Discussion---} In more general states of BCS form, the rotational covariance of $\Delta_\bk$ may be broken, as can spin-rotation symmetry, and even translation symmetry. Because of the chiral form of the low-lying part of the ES (and assuming a local Hamiltonian), the total number of chiral Majorana modes in the ES found here must be robust under such perturbations, unless the system undergoes a phase transition at which the bulk energy gap vanishes. Our approach can also be applied to paired states of other symmetry classes (including those with time-reversal symmetry), to multi-component fermions, and in higher dimensions \cite{rsfl}.

In conclusion, we have shown that for certain model wave-functions for the complex $\ell$-wave superfluids the ES can be computed exactly, and the entanglement gap is infinite. For spinless $\ell$-wave ($\ell$ odd) pairing, the ES consists of $|\ell|$ chiral Majorana fields ($2|\ell|$ for spin singlet). The low--pseudo-energy part of the ES is that of a perturbed chiral CFT which is also that of the edge states \cite{HaldaneLi,ESFQHE,Fidkowski,tzv}, and is expected to be universal.

\acknowledgments

This work was supported by a Yale Postdoctoral Prize Fellowship (JD), and by NSF grant no.\ DMR-1005895 (NR).

\end{document}